\documentclass[12pt]{iopart}

\usepackage{iopams}
\usepackage{setstack}
\usepackage{graphicx}
\usepackage{bm}
\usepackage{verbatim}
\usepackage{epstopdf}
\usepackage{cite}
\DeclareMathAlphabet{\mathpzc}{OT1}{pzc}{m}{it}
\makeatletter
\DeclareRobustCommand{\text}{%
  \ifmmode\expandafter\text@\else\expandafter\mbox\fi}
\let\nfss@text\text
\def\text@#1{{\mathchoice
  {\textdef@\displaystyle\f@size{#1}}%
  {\textdef@\textstyle\f@size{#1}}%
  {\textdef@\textstyle\sf@size{#1}}%
  {\textdef@\textstyle \ssf@size{#1}}%
  \check@mathfonts
  }%
}
\def\textdef@#1#2#3{\hbox{{%
                    \everymath{#1}%
                    \let\f@size#2\selectfont
                    #3}}}
\makeatother

\begin{document}


\title[Abelian cosmic string in the extended Starobinsky model of gravity]{Abelian cosmic string in the extended Starobinsky model of gravity}

\author{J. P. Morais Gra\c ca$^{1}$ and V. B. Bezerra$^{1}$}

\address{$^{1}$ Departamento de F\'{i}sica, Universidade Federal da Para\'{i}ba, Caixa Postal 5008, CEP 58051-970, Jo\~{a}o Pessoa, PB, Brazil}

\ead{jpmorais@gmail.com and valdir@fisica.ufpb.br}

\begin{abstract}
We analyze numerically the behaviour of the solutions corresponding to an Abelian cosmic string taking into account an extension of the Starobinsky model, where the action of general relativity is replaced by $f(R) = R - 2\Lambda + \eta R^2 + \rho R^m$, with $m > 2$. As an interesting result, we find that the angular deficit which characterizes the cosmic string decreases as the parameters $\eta$ and $\rho$ increase. We also find that the cosmic horizon due to the presence of a cosmological constant is affected in such a way that it can grows or shrinks, depending on the vacuum expectation value of the scalar field and on the value of the cosmological constant\end{abstract}

\pacs{04.20.Jb, 04.50.Kd, 04.60.Cf}



\maketitle


%
%
\section{Introduction}
The interest in the Starobinsky model for inflation \cite{Starobinsky:1980te} has been revived, since it appears to be consistent with current cosmological data obtained by the Planck satelite \cite{Planck:2013jfk}. The original model can be recast as an $f(R)$ theory of gravity, where $f(R) = R + \eta R^2$. Another important feature of the Starobinsky model is linked with renormalizability, since the renormalizability of this theory appears to be improved when compared to Einstein's gravity \cite{Stelle:1976gc}. Also, in both supergravity and superstring theories, polynomial corrections to the Einstein-Hilbert action are expected to appear, and so the Starobinsky model can be seen as the first-order correction of a more general effective theory.

Among the many proposals to extend Einstein's gravity, the so-called $f(R)$ theories have deserved some attention also due to the fact that they can model the observed acceleration of the universe without an $ad$ $hoc$ cosmological constant [for a review, see \cite{Sotiriou:2008rp}]. In this context, the Starobinsky model is probably the most natural extension to the Einstein-Hilbert action as an $f(R)$ theory. Thus, it is natural to wonder how a more generic polynomial for $f(R)$, such as $f(R) = R + \eta R^2 + \rho R^m$, $m > 2$, will differ from the Starobinsky model. Inflationary results based on such extended Starobinsky models have been recently studied in \cite{Broy:2014xwa,Broy:2015zba}. Our aim is to study these extended models for cosmic strings.

One of the motivations for the inflationary theory was to explain the current density (or absence) of magnetic monopoles, a heavy and stable particle predicted to be generated at the early universe by mostly Grand Unified Theories (GUTs). Magnetic monopoles are a particular example of a topological defect, a stable solution of classical fields where their stability is provided by topological arguments related to the vacuum manifold. Another example of a topological defect is a comic string (For a review, \cite{Hindmarsh:1994re,Sakellariadou:2009ev}), an axially symmetric relic from phase-transitions ocurred at the early universe. 

If we believe that an effective theory will replace Einstein's gravity, it is natural to study the formation and behaviour of topological defects in the context of it. Despite the fact that the gravitational field far away from the cosmic string is null, it generates an angular deficit that can, in principle, be observed by their astrophysical (and gravitational) effects \cite{Vilenkin:2000jqa}. Until the present moment, no evidence of the existence of cosmic strings has been obtained. As the angular deficit generated by the cosmic string is proportional to the scale of symmetry breaking (the Vacuum Expected Value or VeV), the failure to detect such angular deficit constraints the possible range for the values of the VeV. In \cite{Graca:2015hua} one of us has shown that the Starobinsky model of gravity allows a larger range of values for the VeV than Einstein's gravity, and then the Starobinsky model can be a possible explanation of why we have not observed any evidence of such angular deficit yet.

In this paper we will study the same Abelian Higgs model for cosmic strings considered in \cite{Graca:2015hua}, but now in a more general polynomial $f(R)$ theory. Axially symmetric solutions of more general $f(R)$ theories of gravity have been studied in \cite{Azadi:2008qu,Sharif:2012sv,Momeni:2009tk,Harko:2014axa}, but the results were obtained mainly for generic $f(R)$ functions, or the equations were solved for a fixed energy-momentum tensor. To study the behaviour of the angular deficit and the cosmic horizon as functions of the parameters of the theory, in a dynamical Higgs-Maxwell-Gravity environment, it is necessary to perform a numerical computation. This is done in the present paper. 

One major difference between Starobinsky and extended Starobinsky models are the fact that the former cannot generate or modify a cosmological constant term, but the latter can. We claim that these models should play a fundamental role in the strong curvature regime, such as near compact objects or in the primordial universe era, where it is expected that cosmic strings were generated during some phase transition. This result is independent of the gravitational theory considered. 

The string model we will use was first introduced by Nielsen and Olesen in 1973 as a field-theoretical model to mimic the features of the Nambu-Goto string, at the time a strong candidate to explain Hadronic physics \cite{Nielsen:1973cs}. The gravitational effects of this Abelian Higgs comic string were first analytically studied in \cite{Vilenkin:1981zs} for an idealized cosmic string, and numerically in \cite{Garfinkle:1985hr}. In the later, the Einstein-Maxwell-Higgs equations were solved together, as it should be. Since the coupled equations are highly non-linear, they must be solved numerically. A complete classification of the string-like solutions in Einstein's gravity can be found in \cite{Christensen:1999wb}, and the same cosmic string in the Starobinsky model has been studied in \cite{Graca:2015hua}. Our main goal in this paper is to extend these previous results for a polynomial $f(R)$ theory.

This paper in organized in the following manner. In section 2 we will present our model and the field equations to be solved, and in the following sections they will be solved numerically: In section 3 we will study the asymptotically flat case, and in section 4 we will deal with the asymptotically de Sitter case. Finally, in section 4 we will present the conclusions.

\section{The Model}

The action for the gravitating Abelian Higgs system in this extended Starobinsky model is given by

\begin{equation}
S = \int{d^4x \sqrt{|g|}\left[\frac{1}{2}D_\mu\Phi^* D^\mu \Phi - \frac{\lambda}{4}(\Phi^*\Phi - \nu^2)^2 - \frac{1}{4}F_{\mu\nu}F^{\mu\nu} + \frac{1}{16 \pi G}f(R)\right]}
\label{action}
\end{equation}
where $f(R) = R - 2\Lambda + \eta R^2 + \rho R^m$, with $m \geq 3$. As usual, $F_{\mu\nu}$ is the Abelian field strength, $\Phi$ is a complex scalar field with vacuum expectation value $\nu$, $D_\mu = \nabla_\mu - i e A_\mu$ is the gauge covariant derivative and $\Lambda$ is the cosmological constant. We are using units were $ c = \hbar = 1$. If the parameters $\eta$ and $\rho$ are null, the above action reduces to the Abelian Higgs model in Einstein's gravity with cosmological constant \cite{BezerradeMello:2003ei}

Because of the cylindrical symmetry of the source, and due to the symmetry under boosts along the string axis, we will consider the line element

\begin{equation}
ds^2 = - N^2(r) dt^2 + dr^2 + L^2(r) d\phi^2 + N^2(r) dz^2 ,
\label{metric}
\end{equation}
and the usual Nielsen-Olesen ansatz given by \cite{Nielsen:1973cs}

\begin{eqnarray}
\Phi(r) = \nu f(r) e^{i \phi} 
\\
A_{\mu} dx^\mu = \frac{1}{e} [1 - P(r)] d\phi .
\end{eqnarray}

We chose this ansatz so that the matter fields reach their vacuum expectation values at infinity when we impose the boundary conditions \cite{Nielsen:1973cs}. If we vary the above action with respect to the scalar and gauge fields, respectively, we obtain the following set of field equations

\begin{eqnarray}
\frac{(N^2Lf')'}{N^2L} + (\lambda \nu^2 (1-f^2) - \frac{P^2}{L^2})f = 0
\\
\frac{L}{N^2}(\frac{N^2}{L}P')' - e^2 \nu^2 f^2 P = 0,
\end{eqnarray}
where the symbol ($'$) means derivative with respect to the radial coordinate. The energy-momentum tensor for the matter fields is given by $T^{\mu\nu} = \frac{2}{\sqrt{-g}}\frac{\delta S_{matter}}{\delta g_{\mu\nu}}$, so that its components are

\begin{eqnarray}
& T_t^{\phantom{t}t} = - \epsilon_s - \epsilon_v - \epsilon_w - u \\
\nonumber
& T_r^{\phantom{r}r} = + \epsilon_s + \epsilon_v - \epsilon_w - u \\
\nonumber
& T_\phi^{\phantom{\phi}\phi} = - \epsilon_s + \epsilon_v + \epsilon_w - u \\
\nonumber
& T_z^{\phantom{z}z} = T_t^{\phantom{t}t},
\end{eqnarray}
where 

\begin{eqnarray}
& \epsilon_s = \frac{\nu^2}{2}f'^2,  \hspace{10pt}
\epsilon_v = \frac{P'^2}{2 e^2 L^2}, \hspace{10pt}
\epsilon_w = \frac{\nu^2 P^2 f^2}{2L^2} \hspace{10pt}and\hspace{10pt}
u = \frac{\lambda \nu^4}{4}(1-f^2)^2 .
\end{eqnarray}
The gravitational field equations are obtained varying the action with respect to the metric $g_{\mu\nu}$, and are given by

\begin{equation}
G_{\mu\nu}F(R) + \frac{1}{2} g_{\mu\nu} (2 \Lambda + \eta R^2 + \rho (m-1)R^m) - (\nabla_\mu \nabla_\nu - g_{\mu\nu} \Box) F(R) = \kappa^2 T_{\mu\nu}
\label{eqField1}
\end{equation}
where $F(R) = 1 + 2 \eta R + m \rho R^{m-1}$, $G_{\mu\nu} = R_{\mu\nu} - \frac{1}{2} g_{\mu\nu} R$ and $\kappa^2 = 8 \pi G$. This is a set of three non-zero independent fourth-order differential equations, but to avoid dealing with fourth-order differential equations we will consider the Ricci scalar as an independent field. Taking the trace of the above equations we find

\begin{equation}
-R + 4\Lambda + (m-2)\rho R^m + 3 \Box F(R) = \kappa^2 T .
\label{eq10}
\end{equation}

First, let us point out that Eq. (\ref{eq10}) tells us that the Ricci scalar obeys a differential equation instead of a purely algebraic one, as occurs in general relativity. We can get some information considering a region far away from the string. Assuming that $T = 0$ and the cosmological constant vanishes, a constant curvature solution is given by

\begin{equation}
-R + (m-2) \rho R^m = 0,
\label{solutionR}
\end{equation}
which means that we can, in principle, obtain a constant non-zero solution for the scalar curvature $R$, if $m \geq 3$. This constant, however, will be proportional to the inverse of $\rho$. Therefore, a small $\rho$ will result in a large $R$. To be able to work with a small cosmological constant, we will insert the $\Lambda$ parameter. For $m = 2$, the unique solution is the asymptotically flat case. 

Inserting Eq. (\ref{eq10}) into Eq. (\ref{eqField1}), we get

\begin{equation}
 \hspace*{-60px} F(R)G_{\mu\nu} + g_{\mu\nu}\left(\frac{R}{3} + \eta\frac{R^2}{2} + \rho \frac{m+1}{6} R^m - \frac{\Lambda}{3}\right) - \nabla_{\mu} \nabla_{\nu}F(R)  = \kappa^2 \left(T_{\mu\nu} - \frac{g_{\mu\nu}T}{3}\right) .
\end{equation}

Before we deal with the components of the above differential equations, we should redefine some variables as well as field, in such a way that all parameters are dimensionless. To achieve this goal we will express all lengths in terms of the scalar characteristic length scale given by $1/\sqrt{\lambda \nu^2}$. We will change our radial coordinate to a dimensionless coordinate $x = \sqrt{\lambda \nu^2} r$, together with the field redefinitions $L(x) = \sqrt{\lambda \nu^2} L(r)$ and $R(x) = R(r) / \lambda \nu^2$. We also introduce four new parameters, $\alpha = e^2 / \lambda $, $\gamma = 8 \pi G^2 \nu^2 $, $ \xi = \eta \lambda \nu^2 $ and $\chi = \rho (\lambda \nu^2)^{m-1}$. Thus, we can write the field equations as

\begin{eqnarray}
&f'' = (\frac{P^2}{L^2} + f^2 - 1)f - \frac{(N^2L)'f'}{N^2L} ,
\label{eqqF}
\\
&P'' = \alpha f^2 P - \frac{L}{N^2}(\frac{N^2}{L})'P' ,
\label{eqqP}
\end{eqnarray}
for the matter fields and

\begin{eqnarray}
\label{ch3:eqR}
R'' = \frac{1}{12} \frac{1}{N^2 L^2 \alpha R (2 \xi R^2 + m(m-1) \chi R^m)} [ 6 \xi \alpha L R^4 (R N^2 L + 4 N'^2  L  
\\
\nonumber
+ 8 N' N L') + 2 \chi \alpha R^m L ((m+1) R^3 N^2 L - 6 m(m^2 - 3m + 2) R'^2 N^2 L
\\
\nonumber
 + 6 m R^2 N'^2 L + 12 m R^2 N' N L') -\gamma R^3 N^2(10 \alpha f'^2 L^2 - 2 \alpha P^2 f^2 
\\
\nonumber 
+ \alpha L^2(1 - f^2)^2+ 6 P'^2 ) + 4 \alpha R^2 L (6 N' N L' + 3 N'^2 L + R N^2 L)
\\
\nonumber
- 4\Lambda \alpha N^2 L^2 R^3],
\end{eqnarray}

\begin{eqnarray}
\label{ch3:eqN}
N'' = \frac{-1}{24} \frac{1}{\alpha L^2 N(R + 2 \xi R^2 + m \chi R^m)}[ 6 \xi \alpha R L (4 R L N'^2 + R^2 L N^2  \\
\nonumber
- 4 R' L' N^2) + 2 \chi \alpha R^2 L ((m+1) R^{m-1} N^2 L - 6m(m-1) R' L' N^2 \\
\nonumber
+ 6m L R^{m-2} N'^2) - \gamma R N^2(10 \alpha P^2 f^2 - 2 \alpha f'^2 L^2 + \alpha L^2(1 - f^2)^2+ 6 P'^2 ) \\
\nonumber
+ 4 \alpha R L^2 (N^2 R + 3 N'^2) - 4 \Lambda \alpha N^2 L^2 R],
\end{eqnarray}

\begin{eqnarray}
\label{ch3:eqL23}
L'' = \frac{1}{24} \frac{1}{\alpha L N^2(1 + 2 \xi R + m \chi R^{m-1})}[6 \xi \alpha L(4 N'^2 L R - R^2 N^2 L  \\
\nonumber
- 8 N' N L' R + 8 R' N' L N - 4 R' L' N^2) - 2 \alpha \chi R L (2 (m+1) R^{m-1} N^2 L \\
\nonumber
- 6m R^{m-2} L N'^2 + 12 m N' R^{m-2} N L' - 12 m(m-1) R' N' L N R^{m-3} \\
\nonumber
+ 6m(m-1) R^{m-3} N^2 R' L') - \gamma N^2(\alpha (2 f'^2 L^2 + 14 P^2 f^2 - L^2 (1- f^2)^2) \\   
\nonumber
+ 18 P'^2) + 4 \alpha L(3 N'^2 L - R N^2 L - 6 N' N L') + 4 \Lambda \alpha L^2 N^2].
\end{eqnarray}
for the metric functions, where the primes $(')$ in the above equations mean derivatives with respect to the new coordinate $x$. Equations (\ref{eqqF}) and (\ref{eqqP}) give us the dynamics of the matter fields, while Eqs (\ref{ch3:eqR}) to (\ref{ch3:eqL23}) give us the behaviour of the metric fields as well as of the Ricci scalar field. We must also impose the boundary conditions for all functions. As mentioned, in order to get string-like solutions the matter fields should reach their vacuum expectation values asymptotically. We must then impose the following boundary conditions

\begin{eqnarray}
f(0) = 0, \hspace{10pt} f(\infty) = 1, \\
\nonumber
P(0) = 1, \hspace{10pt} P(\infty) = 0,
\end{eqnarray}
where infinity should be understood as a limit. The boundary conditions at the origin are necessary to have regular fields. 

On the other hand, the boundary conditions for the metric functions are given by

\begin{eqnarray}
L(0) = 0, \hspace{10pt} L'(0) = 1, \\
\nonumber
N(0) = 1, \hspace{10pt} N'(0) = 0, \\
\end{eqnarray}
which were imposed to guarantee the regularity of the metric. We must now impose the boundary conditions for the Ricci scalar field. We are looking for asymptotically constant Ricci spacetime, such as (anti-)de Sitter or Minkowski. Then, we will impose the following boundary conditions

\begin{eqnarray}
R(\infty) = R_{c}, \hspace{10pt} R'(\infty) = 0, \\
\end{eqnarray}
where $R_{c}$ is a constant. In this case, we have two kinds of solutions. The first one is $R_{c} = 0$, as long as we also impose $\Lambda = 0$. This branch, which is asymptotically flat, will be studied in Section 3. The second branch is given by $R_{c}$ as a solution of Eq. (\ref{solutionR}) with a cosmological constant. Considering a small value for $\Lambda$, we can write the constant Ricci scalar as $R_{c} = 4\Lambda + \epsilon$, since $\rho R_{c}^3$ will be much smaller (at least for reasonable values of $\rho$). Introducing this ansatz in Eq. (\ref{solutionR}) and neglecting second-order terms in $\epsilon$, we have

\begin{equation}
-\epsilon + (m-2)\rho (4\Lambda)^m + m \epsilon (4 \Lambda)^{m-1} = 0,
\end{equation}
which gives us the boundary condition

\begin{equation}
R(\infty) = 4\Lambda + \frac{(m-2)\rho (4\Lambda)^m}{1 - m \rho (4\Lambda)^{m-1}}.
\end{equation}

We now have to solve a four parameter set of five differential equations with boundary conditions. As we can see, these equations are highly non-linear and will be solved numerically. The behaviour of such string was already studied in general relativity without a cosmological constant \cite{Garfinkle:1985hr}, with a cosmological constant \cite{BezerradeMello:2003ei} and in the Starobinsky model of gravity \cite{Graca:2015hua}. Our aim is to enlarge this study including more terms on a polynomial expansion with a cosmological constant.

\subsection{The conical geometry}

In an asymptotically flat spacetime, the geometry induced by the cosmic string far away from its core is a flat conical geometry. Asymptotically, the metric functions are given by

\begin{eqnarray}
N(x \rightarrow \infty) = a,
\\
L(x \rightarrow \infty) = bx + c, \hspace{10pt} b \geq 0,
\end{eqnarray} 
where $a$, $b$ and $c$ are constants depending on $\alpha, \gamma$ and any other gravitational parameters. In our model, there are two parameters, namely $\xi$ and $\chi$. In the absence of any source it is expected that $a=1, b=1$ and $c=0$. 

The parameter $\alpha$ is proportional to the relation between the scalar and gauge masses, and does not strongly influences the asymptotically angular deficit. It is the parameter $\gamma$ that considerably affects the space-time topology. In fact, if we fix $\alpha$, $\xi$ and $\chi$, as we vary $\gamma$ we are in a way increasing the coupling between the matter fields and the gravitational field. As mentioned, the string-like solution gives rise to a planar deficit angle that can be expressed as

\begin{equation}
\Delta = 2\pi(1- L'(\infty)),
\label{eqDelta}
\end{equation}
where $L'(\infty)$ is inversely proportional to $\gamma$. It is worth calling attention to the fact that we can increase $\gamma$ until $\Delta$ reaches $2\pi$, which means that $L'(\infty) = 0$. This value is called critical $\gamma$, $\gamma_{cr}$, and for $\gamma > \gamma_{cr}$ the spacetime is not globally well-defined in some models. Let us remember that $\gamma$ is directly related to the scale of the symmetry breaking. We can then use $\gamma_{cr}$ as a kind of constraint on the maximum scale for the occurrence of it. In a real world scenario, we should expect only a small value for the angular deficit to be allowed, otherwise its effects should already be noted. To deal with critical values, however, is useful for a better understanding of the properties of the theory.

In general relativity, which are given by $\xi = 0$, we can find the critical values for $\gamma$ as a function of $\alpha$ \cite{Brihaye:2000qr}, 

\begin{equation}
\gamma_{cr}(1.0) \approx	1.66, \hspace{10pt} \gamma_{cr}(3.0) \approx 2.2, \hspace{10pt} \text{and so on}.
\end{equation}

In \cite{Graca:2015hua}, one of us studied how this model is affected as we increase $\xi$. The obtained result tell us that, for a fixed set of ($\alpha$, $\gamma$), as $\xi$ gets bigger, the angular deficit becomes smaller. This means that the Starobinsky model of gravity allows a large range for the values of the symmetry-breaking scale (or VeV). The angular deficit generated by a large VeV will result, in the Starobinsky model of gravity, in a smaller angular deficit than in general relativity. We can conclude that the failure to observe a conical geometry can be due to corrections on general relativity other than due to a small VeV.

Our aim in the next two sections is to study how these results change as we extend the Starobinsky model, including more terms in a polynomial expansion. 

\subsection{The cosmological horizon}

The inclusion of a cosmological constant in the gravitational equations gives rise to a cosmological horizon, beyond which all events are not causally connected. In an axially symmetric space-time in vacuum, the horizon can be obtained by the line element (\ref{metric}) with the metric functions given by

\begin{equation}
N(x) = cos\left(\frac{\sqrt{3 \Lambda}x}{2}\right)^{3/2}
\label{ch3:NdeSitter}
\end{equation}
and
\begin{equation}
L(x) = \frac{2^{2/3}}{\sqrt{3\Lambda}} \left[ sin(\sqrt{3\Lambda}x)\right]^{1/3} \left[ tan(\sqrt{3\Lambda} \frac{x}{2})\right]^{2/3}.
\end{equation}
This solution can be seen as the space-time generated by the cosmic string when $\gamma = 0$. In this regime, the matter fields decouple from the gravitational fields. The cosmological horizon is then given by the first zero of $N(x)$, which means $x_{ch} = \pi / \sqrt{3 \Lambda}$. In general relativity, as we increase $\gamma$, the value of $x_{ch}$ decreases. We will study how this result changes as we move from general relativity to the extended Starobinsky model.

\section{Asymptotically flat spacetime}

We have used a finite difference Newton-Raphson algorithm with adaptive grid scheme \cite{Archer1}\cite{Archer2} to construct the solutions numerically. Our estimated errors range from $10^{-8}$ to $10^{-12}$, sometimes even better. The limit $\xi \rightarrow 0$, $\chi \rightarrow 0$ corresponds to general relativity, but we cannot use both $\xi = 0$ and $\chi = 0$ in our analysis because the equations are not well defined at these values. The reason is that, in Einstein's gravity, the Ricci scalar obeys an algebraic relation with the energy-momentum tensor, not a differential equation. 

\begin{figure}[htb]
\centering
\begin{tabular}{@{}cc@{}}
\includegraphics[scale=0.8]{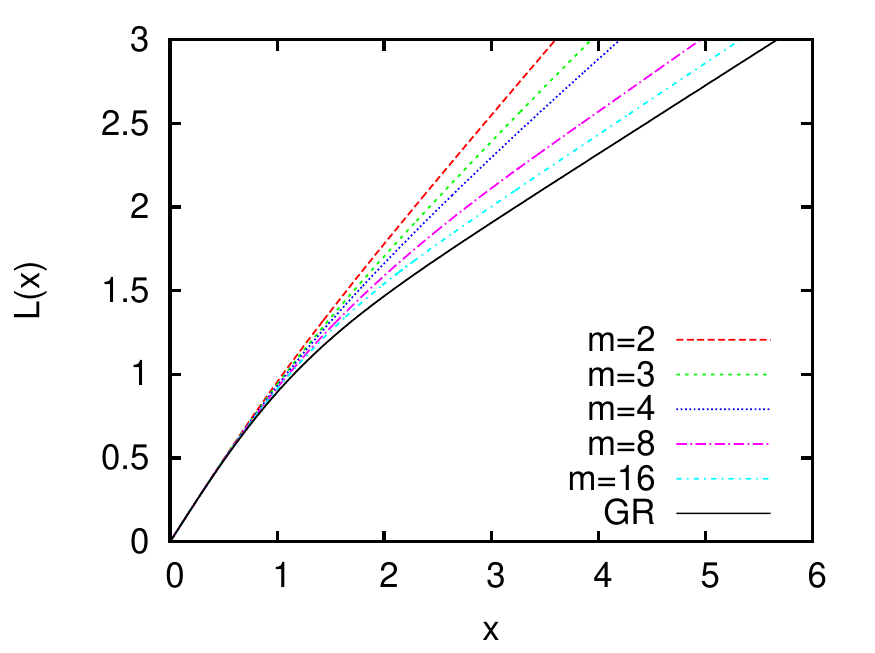} &
\includegraphics[scale=0.8]{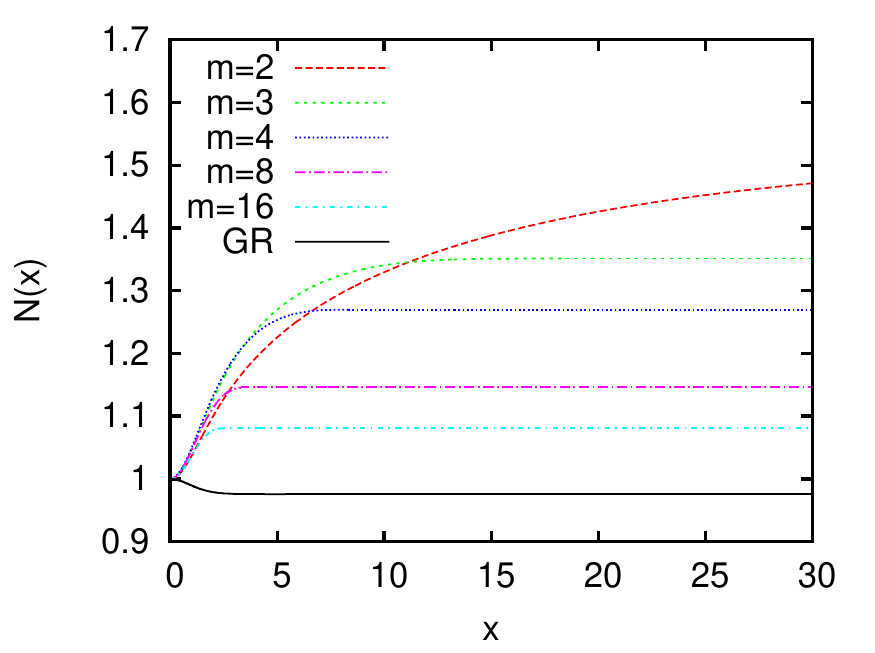}
\end{tabular}
\caption{The metric functions for $\alpha = 1.0$, $\gamma = 1.0$, $\xi = 0.01$ and $\chi = 100$ in the extended Starobinsky model. The continuous line represents the same system in Einstein's gravity.}
\label{CompareLN}
\end{figure}

In \cite{Graca:2015hua} it has been shown that, for $\chi = 0$, the angular deficit decreases as the parameter $\xi$ increases. From Eq. (\ref{eqDelta}) we see that the angular deficit is null for $L'(\infty) = 1$ and reaches $2 \pi$ for $L'(\infty) = 0$. This means that we can measure the angular deficit by the inclination of the $L(x)$ metric function. In Figure (\ref{CompareLN}) we plot $L(x)$ and $N(x)$ metric functions with $\alpha = 1.0$, $\gamma = 1.0$, $\xi = 0.01$ and $\chi = 100$, for $m=2,3,4,8$ and $16$. The plots compare these functions with the metric function in general relativity. It can be seen that, as the power of the polynomial increases, the correction relative to general relativity becomes smaller. We are led to expect that, for $m \rightarrow \infty$, the model approaches general relativity.

Since we expect that any well-behaved $f(R)$ theory can be expanded in a power series, we are led to expect that only the first few terms in the power series have some relevance. This claim can be avoided if, for some reason, the coefficients in the series increase exponentially as we increase the power $m$ of the polynomial. However, we see no reason for this behaviour.

The metric function $N(x)$ is also ploted in Figure (\ref{CompareLN}). It also approaches general relativity as we increase the polynomial power $m$. For an asymptotically flat space-time, this $N(x)$ function do not significantly influences the angular deficit, but do influences the energy of the string. 

\begin{figure}[htb]
\centering
\begin{tabular}{@{}cc@{}}
\includegraphics[scale=0.8]{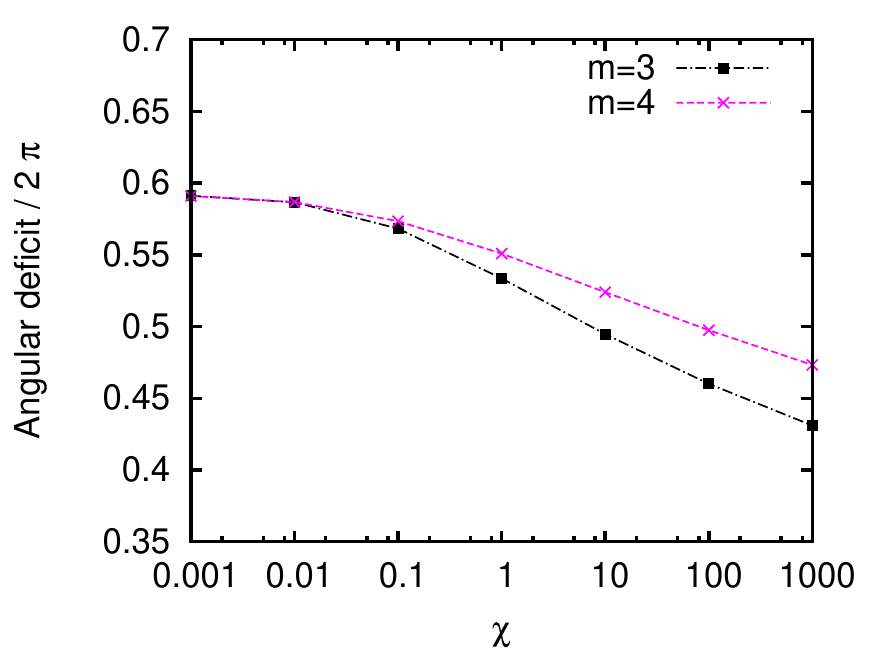} &
\includegraphics[scale=0.8]{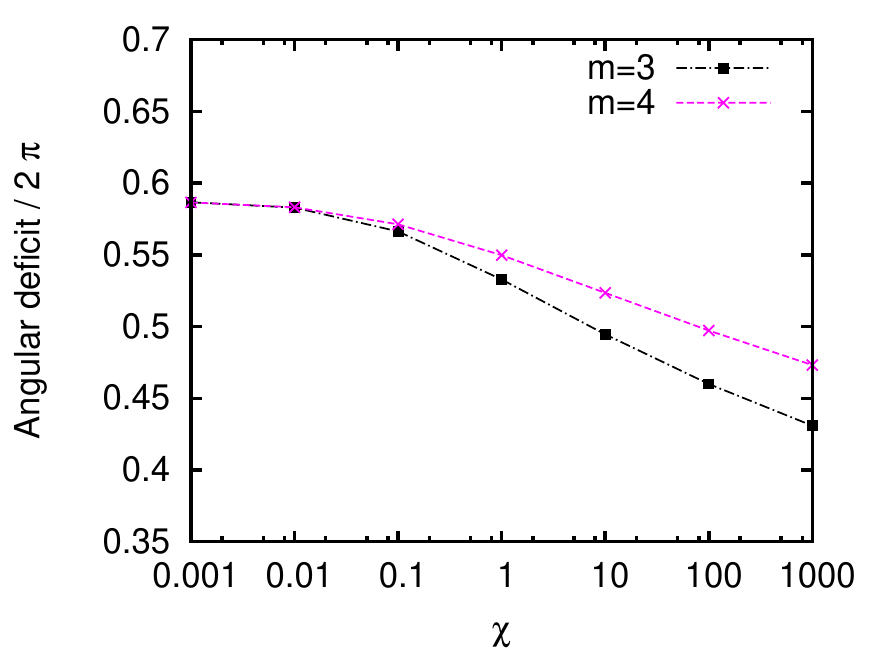} \\
\includegraphics[scale=0.8]{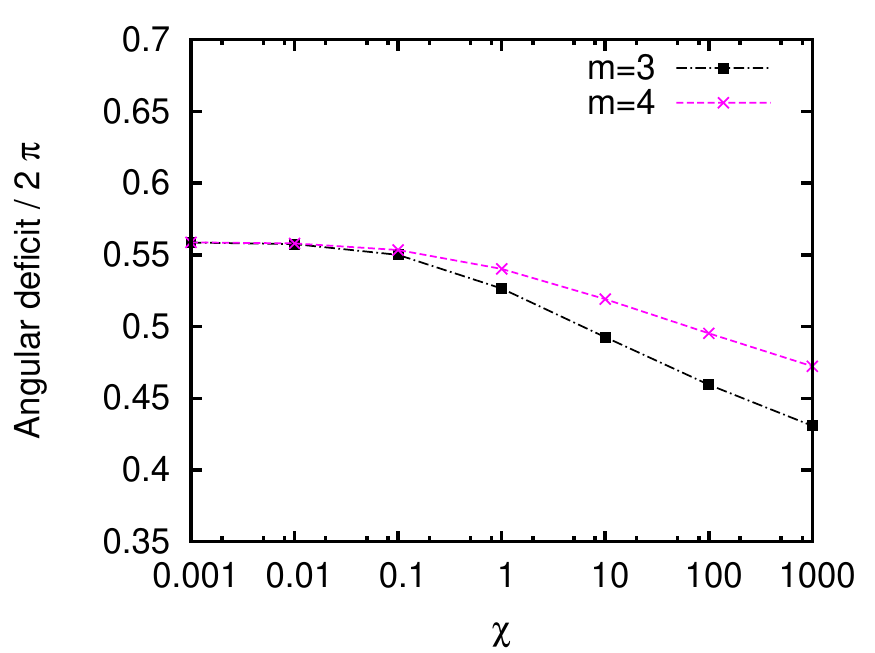} &
\includegraphics[scale=0.8]{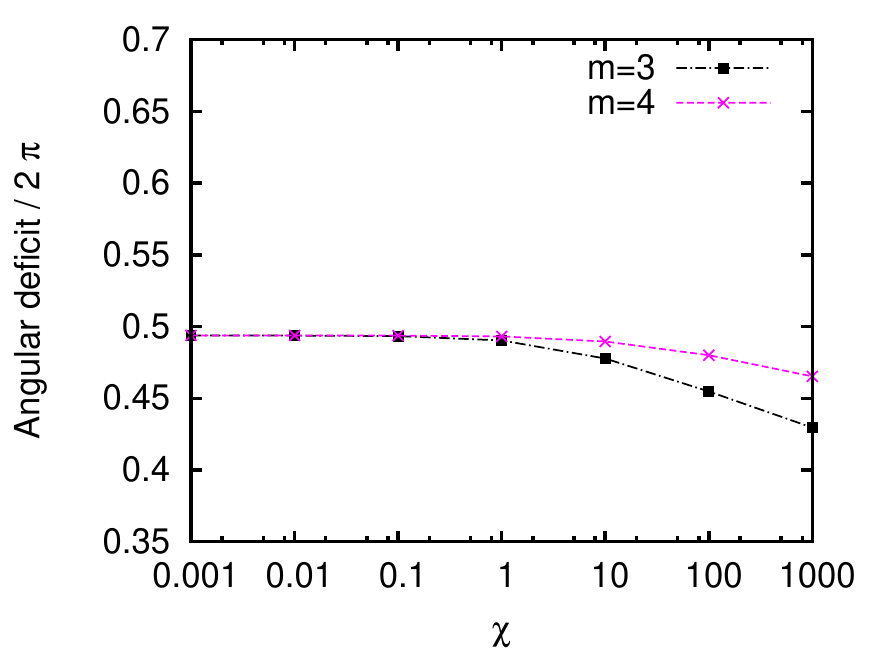} \\
\includegraphics[scale=0.8]{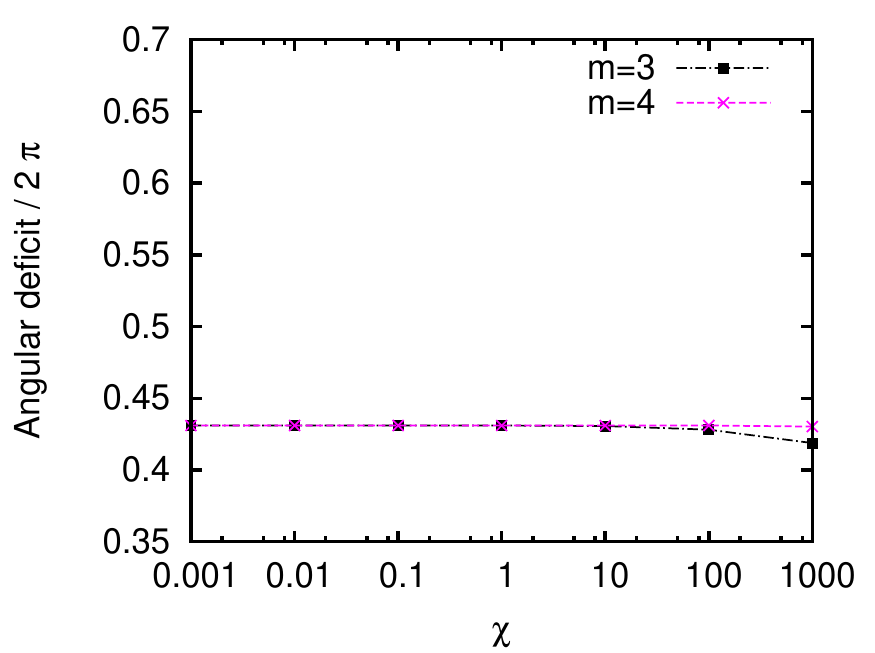} &
\includegraphics[scale=0.8]{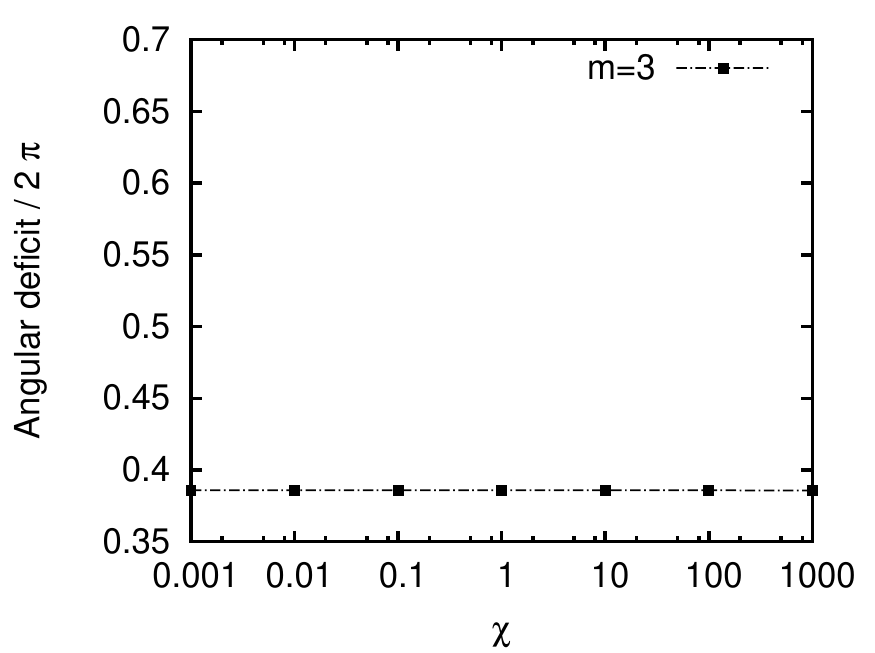}
\end{tabular}
\caption{The angular deficit over $2 \pi$ generate by the Abelian Higgs model. From the upper left plot to the right, the parameter $\xi$ is given by $0.001$, $0.01$, $0.1$, $1$, $10$ and $100$.}
\label{CriticalGammaExtended}
\end{figure}

In Figure (\ref{CriticalGammaExtended}) it is plotted how the angular deficit changes as we vary $\xi$ and $\chi$. As we have two parameters, each individual plot was calculated for a fixed value for $\xi$. To see the effect of the extended Starobinsky model, we then vary the parameter $\chi$ from $0.001$ to $1000$. This procedure was done for $m=3$ and $m=4$ and it is expected that a similar behaviour would be obtained for $m > 4$. The six plots, from the left upper one to the bottom right, were calculated for $\xi = 0.001, 0.01, 0.1, 1, 10$ and $100$, respectivelly. The obtained result is that the angular deficit become smaller as we increase both $\xi$ and $\chi$, but the effect of "shrinking" the angular deficit is more effective as lower the value of the power $m$. This means that the pure Starobinsky term $R^2$ is the leading-order term. For $\xi = 100$ and $\chi = 0$, the angular deficit is about $0.39$. But for $\xi = 0.01$ and $\chi = 1000$, the angular deficit is about $0.44$ and $0.47$ for $m=3$ and $m=4$, respectivelly.

This means that, as we move from the Starobinsky model to the extended Starobinsky model, the angular deficit continues to shrink, but the correction becomes less and less relevant as we increase the power $m$ of the polynomial. In the left bottom plot ($\xi = 10$), it is possible to see that even for $\chi = 1000$, the $m=4$ polynomial term do not affects anymore the angular deficit, and the $m=3$ polynomial term barely affects it.

\begin{figure}[htb]
\centering
\begin{tabular}{@{}cc@{}}
\includegraphics[scale=0.8]{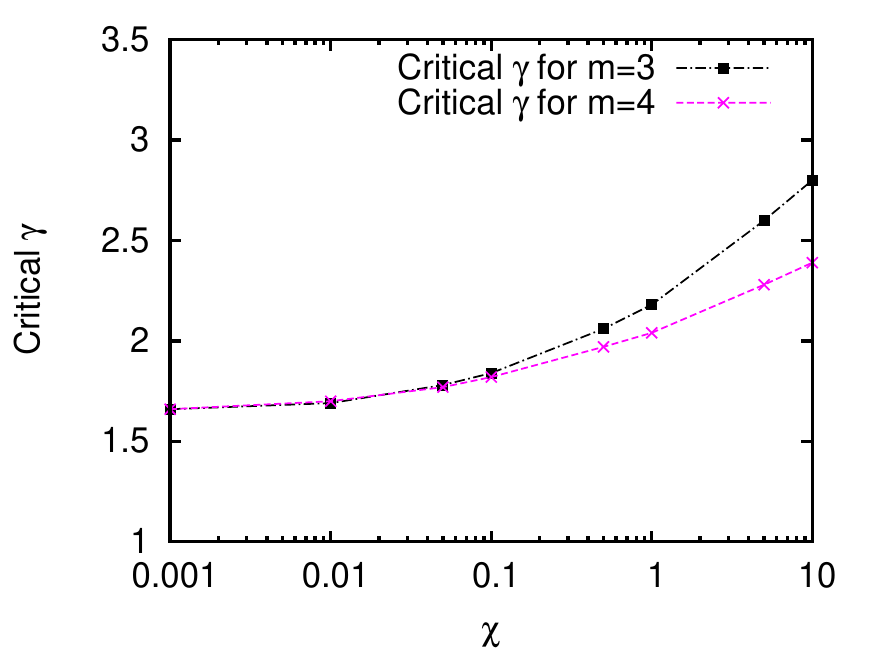} &
\includegraphics[scale=0.8]{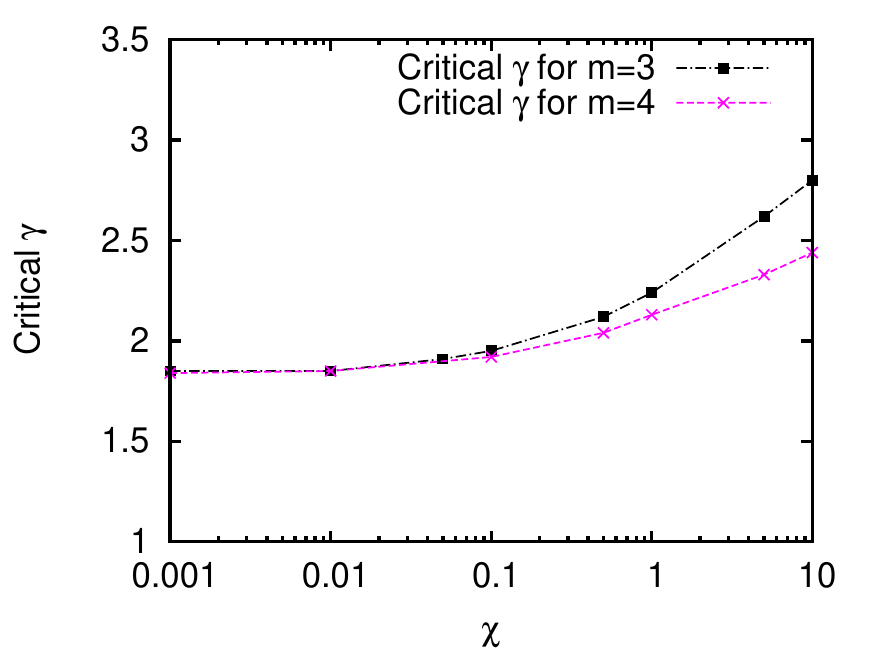} \\
\includegraphics[scale=0.8]{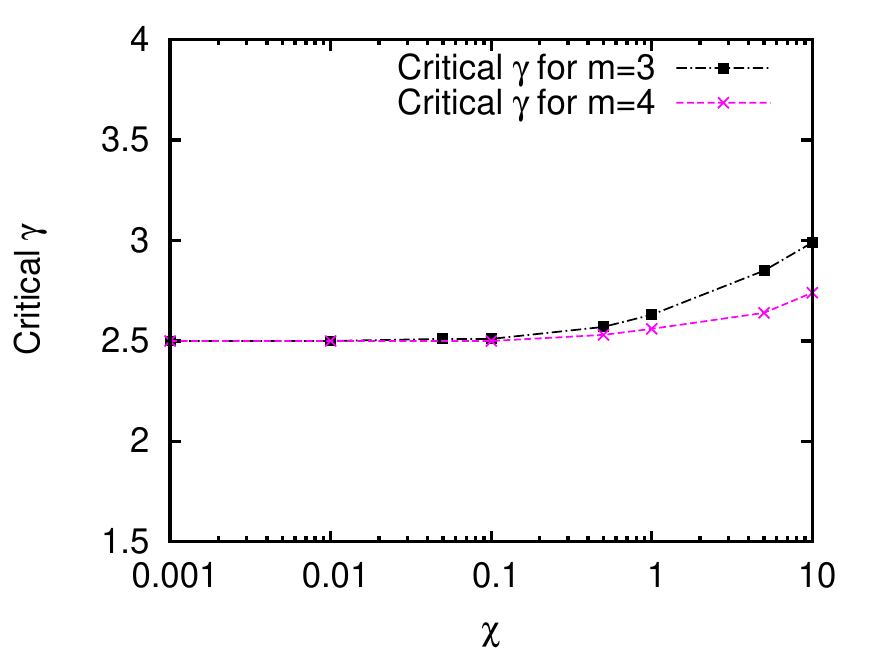} &
\includegraphics[scale=0.8]{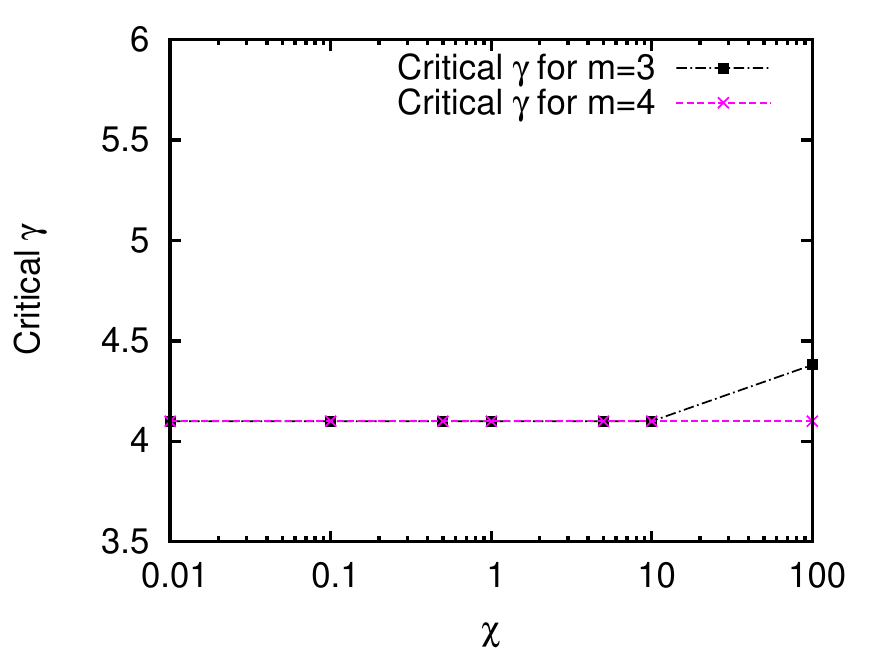}
\end{tabular}
\caption{The value for $\gamma_{cr}$ as an upper bound on the maximum VeV for the scalar field. From the upper left plot to the right, we have, respectively, $\xi = 0.001, 0.1, 1$ and $10$.}
\label{CriticalGammaExtended2}
\end{figure}

The same results can be seen in Figure (\ref{CriticalGammaExtended2}). These fours plots measure how the critical $\gamma$ change as we vary both $\xi$ and $\chi$. Let us remember that $\gamma_{cr}$ is the value when $L(\infty) \rightarrow 0$, which means that the angular deficit reaches $2 \pi$ and this can be considered as an upper bound for the value for $\gamma$. From the upper left plot to the right, we have, respectively, $\xi = 0.001, 0.1, 1$ and $10$. Since $\gamma_{cr}$ is an upper bound, all the values below the lines are acceptable values for $\gamma$. As $\gamma$ is related to the scale of the symmetry breaking, it is an upper bound on the acceptable values for the VeV. What we can see in Figure (\ref{CriticalGammaExtended2}) is that, as we increase both $\xi$ and $\chi$ the upper bound increases, and this is more evident for $m = 3$ than for $m = 4$. In fact, it increases more for the Starobinsky model, since it corresponds to $m=2$.

In a real world scenario, we do not expect $\gamma_{cr}$ to be the upper constraint on the scale of symmetry breaking, but the pattern will be the same. In \cite{Graca:2015hua} it was already been shown that the Starobinsky model allows the scalar field to assume a VeV larger than it would be possible in Einstein's gravity, assuming observational constraints. What has been shown in this section is that the extended Starobinsky increases this effect, but in a less extend than the original Starobinsky model. For a value close to unity for the parameter $\chi$, only the first terms in a power series will contribute to the upper bound on $\gamma_{cr}$.

The way the matter fields are affected in the Starobinsky model of gravity was also already studied in \cite{Graca:2015hua}. It has been shown that the matter preserves its string-like structure and their shape is just slightly changed. The same occurs in this extended Starobinsky model, and we will just concern with the energy of the string, since it can be responsible for gravitational effects. We will follow \cite{Brihaye:2000qr} and define the energy (mass per unit length) as

\begin{eqnarray}
G \mathcal{M} = \frac{\gamma}{8} M_{in} = \frac{\gamma}{8} \int^{\infty}_{0} dx NL \left( (f')^2 + \frac{(P')^2}{\alpha L^2} + \frac{P^2 f^2}{L^2} + \frac{1}{2}(1-f^2)^2 \right).
\label{eqnEnergy}
\end{eqnarray}

With this definition, energy and angular deficit for several values for $\xi$ and $\chi$ were calculated. The results for $m=3$ and $m=4$ are given in Tables \ref{energiaDeficitM3} and \ref{energiaDeficitM4}, respectivelly.

\begin{table}[]
\centering
\caption{Energy (as defined in Eq. \ref{eqnEnergy}) and angular deficit for $\alpha = 1.0$, $\gamma = 1.0$ and $m = 3$.}
\label{energiaDeficitM3}
\begin{tabular}{p{0.2\textwidth}p{0.2\textwidth}p{0.2\textwidth}p{0.2\textwidth}}
\hline
 $\xi$  & $\chi$  & $M_{in}$ & $\delta \phi / 2\pi$  \\ \hline
 0.001  & 0  & 1.1710 & 0.5918 \\
 0.001  & 0.001  & 1.1739 & 0.5911 \\
 0.001  & 0.01  & 1.1867 & 0.5867  \\
 0.001  & 0.1  & 1.2130 & 0.5683  \\
 0.001  & 1.0  & 1.2329 & 0.5335  \\
 0.001  & 10  & 1.2372 & 0.4947  \\
 0.001  & 100  & 1.2331 & 0.4600  \\
 0.001  & 1000  & 1.2266 & 0.4310  \\  
 0.01   & 0  & 1.2266 & 0.5872  \\
 0.01   & 0.001  & 1.1800 & 0.5867  \\
 0.01   & 0.01  & 1.1899 & 0.5829  \\
 0.01   & 0.1  & 1.2136 & 0.5663  \\
 0.01   & 1  & 1.2329 & 0.5328  \\
 0.01   & 10  & 1.2372 & 0.4945  \\
 0.01   & 100  & 1.2331 & 0.4599  \\
 0.01   & 1000  & 1.2266 & 0.4309  \\
 0.1    & 0.001  & 1.2065 & 0.5586  \\
 0.1    & 0.01  & 1.2083 & 0.5575  \\
 0.1    & 0.1  & 1.2187 & 0.5500  \\
 0.1    & 1  & 1.2328 & 0.5265  \\
 0.1    & 10  & 1.2369 & 0.4926  \\
 0.1    & 100  & 1.2329 & 0.4594  \\
 0.1    & 1000  & 1.2266 & 0.4308  \\
 1    & 0.001  & 1.2307 & 0.4938  \\
 1    & 0.01  & 1.2307 & 0.4937  \\
 1    & 0.1  & 1.2310 & 0.4933  \\
 1    & 1  & 1.2325 & 0.4903  \\
 1    & 10  & 1.2345 & 0.4777  \\
 1    & 100  & 1.2319 & 0.4547  \\
 1    & 1000  & 1.2262 & 0.4295  \\ 
 10   & 0.1  & 1.2262 & 0.4310  \\
 10   & 10  & 1.2262 & 0.4306  \\
 10   & 100  & 1.2259 & 0.4281  \\
 10   & 1000  & 1.2237 & 0.4188  \\
 100  & 1  & 1.2153 & 0.3858  \\
 100  & 100 & 1.2153 & 0.3858 \\
 100  & 1000 & 1.2152 & 0.3856
 \end{tabular}
\end{table}

\begin{table}[]
\centering
\caption{Energy (as defined in Eq. \ref{eqnEnergy}) and angular deficit for $\alpha = 1.0$, $\gamma = 1.0$ and $m = 4$.}
\label{energiaDeficitM4}
\begin{tabular}{p{0.2\textwidth}p{0.2\textwidth}p{0.2\textwidth}p{0.2\textwidth}}
\hline
 $\xi$  & $\chi$  & $M_{in}$ & $\delta \phi / 2\pi$  \\ \hline
 0.001  & 0.001  & 1.1766 & 0.5910 \\
 0.001  & 0.01  & 1.1916 & 0.5868  \\
 0.001  & 0.1  & 1.2138 & 0.5735  \\
 0.001  & 1.0  & 1.2311 & 0.5509  \\
 0.001  & 10  & 1.2387 & 0.5240  \\
 0.001  & 100  & 1.2382 & 0.4974  \\
 0.001  & 1000  & 1.2362 & 0.4732  \\  
 0.01   & 0.001  & 1.1816 & 0.5866  \\
 0.01   & 0.01  & 1.1948 & 0.5832  \\
 0.01   & 0.1  & 1.2142 & 0.5713  \\
 0.01   & 1  & 1.2311 & 0.5498  \\
 0.01   & 10  & 1.2372 & 0.5235  \\
 0.01   & 100  & 1.2392 & 0.4972  \\
 0.01   & 1000  & 1.2361 & 0.4731  \\
 0.1    & 0.001  & 1.2066 & 0.5587  \\
 0.1    & 0.01  & 1.2089 & 0.5580  \\
 0.1    & 0.1  & 1.2181 & 0.5534  \\
 0.1    & 1  & 1.2307 & 0.5402  \\
 0.1    & 10  & 1.2380 & 0.5191  \\
 0.1    & 100  & 1.2388 & 0.4953  \\
 0.1    & 1000  & 1.2360 & 0.4723  \\
 1    & 0.001  & 1.2307 & 0.4938  \\
 1    & 0.01  & 1.2307 & 0.4938  \\
 1    & 0.1  & 1.2308 & 0.4937  \\
 1    & 1  & 1.2316 & 0.4931  \\
 1    & 10  & 1.2304 & 0.4896  \\
 1    & 100  & 1.2357 & 0.4800  \\
 1    & 1000  & 1.2344 & 0.4652  \\ 
 10   & 0.1  & 1.2262 & 0.4310  \\
 10   & 10  & 1.2262 & 0.4310  \\
 10   & 100  & 1.2262 & 0.4310  \\
 10   & 1000  & 1.2263 & 0.4302  \\
\end{tabular}
\end{table}

\section{Asymptotically de Sitter space}

In an asymptotically de Sitter space, the question that will concern us is the location of the cosmic horizon. We must study the original Starobinsky model and its extended version separately, since the former gives rise to a cosmological constant term, and the latter does not. As we will see, this difference will characterize how the cosmic horizon, denoted by $x_{ch}$, change as we increase the parameters $\xi$ and $\chi$, related to the gravitational theory. In general relativity with a cosmological constant term, the cosmic horizon is a function of the parameters $\gamma$ and $\Lambda$. It become smaller as both parameters increases.

\begin{figure}[htb]
\centering
\begin{tabular}{@{}cc@{}}
\includegraphics[scale=0.8]{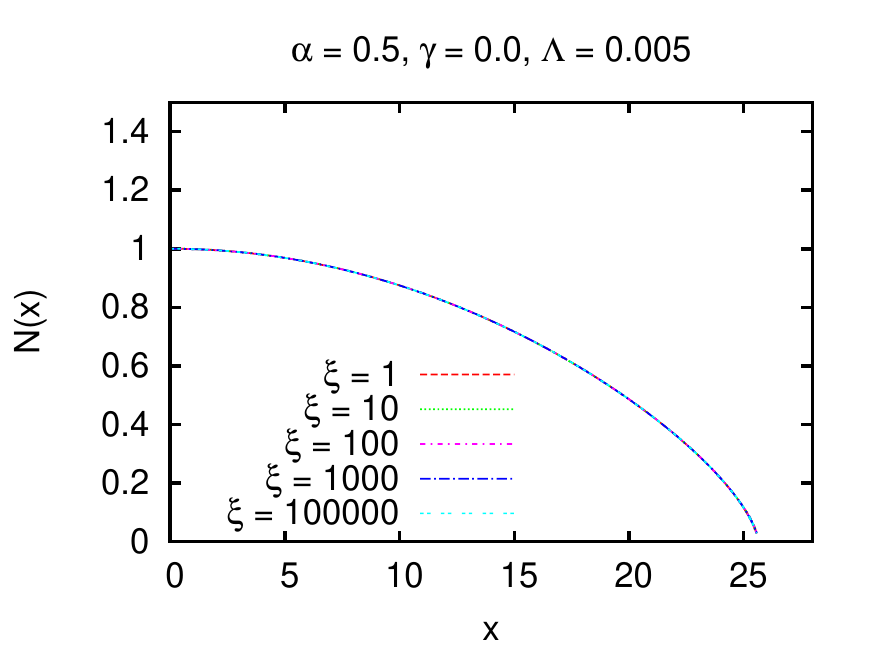} &
\includegraphics[scale=0.8] {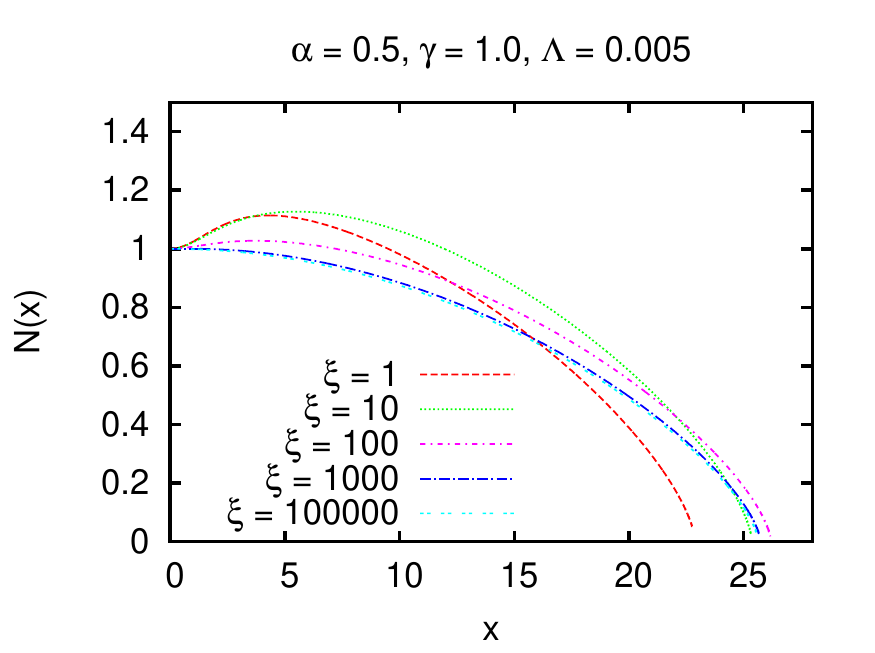} \\
\includegraphics[scale=0.8]{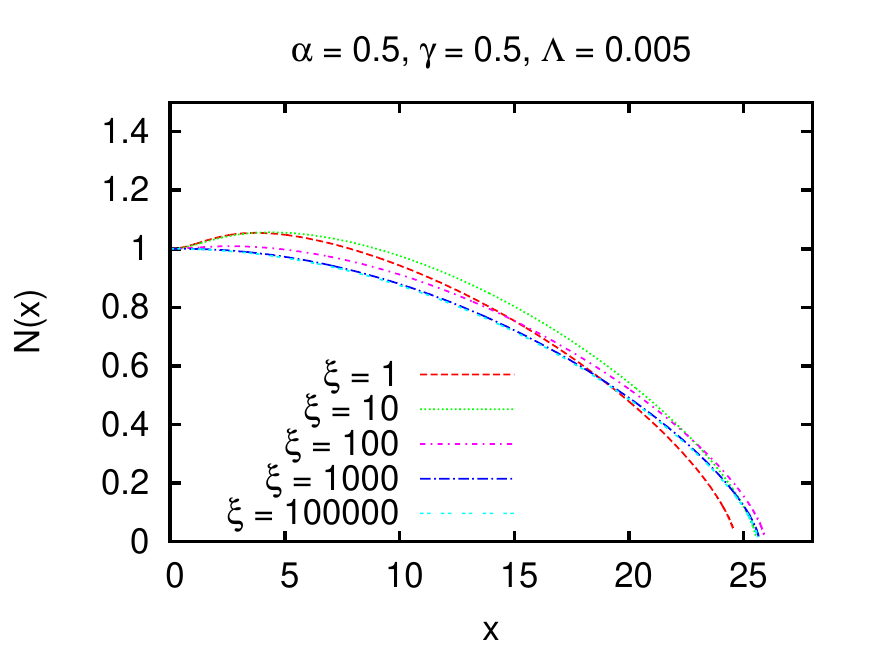} &
\includegraphics[scale=0.8] {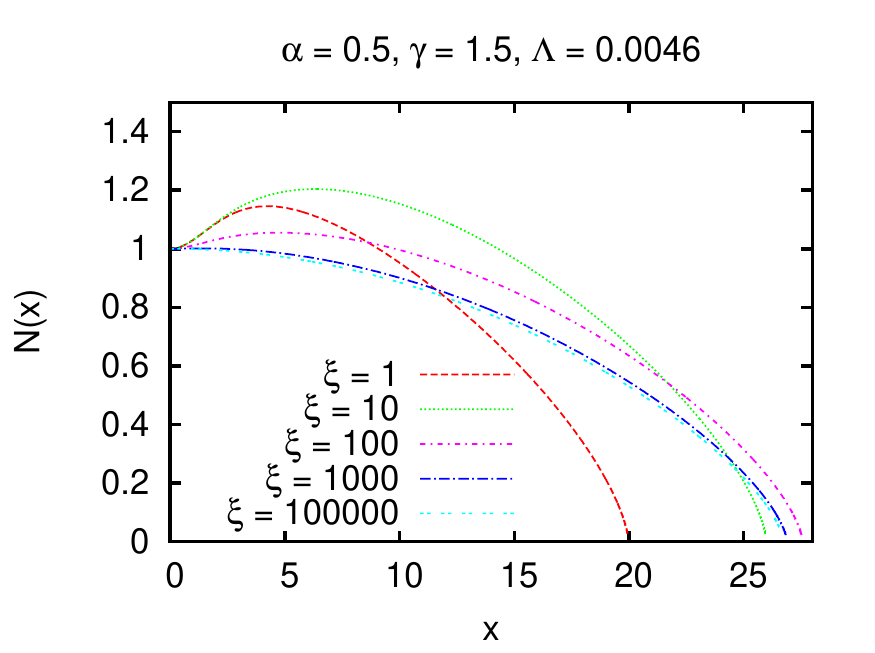} \\
\includegraphics[scale=0.8]{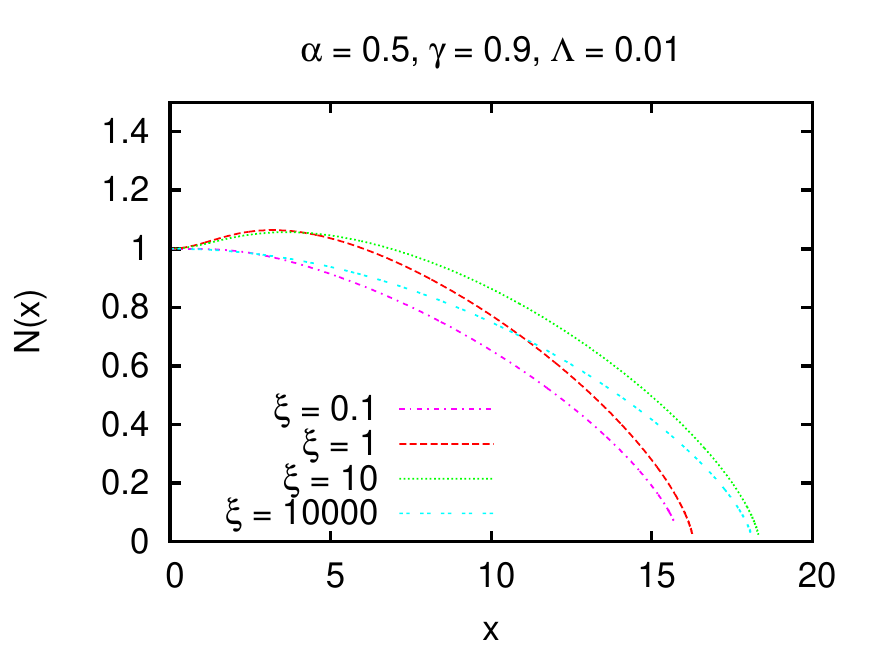} &
\includegraphics[scale=0.8] {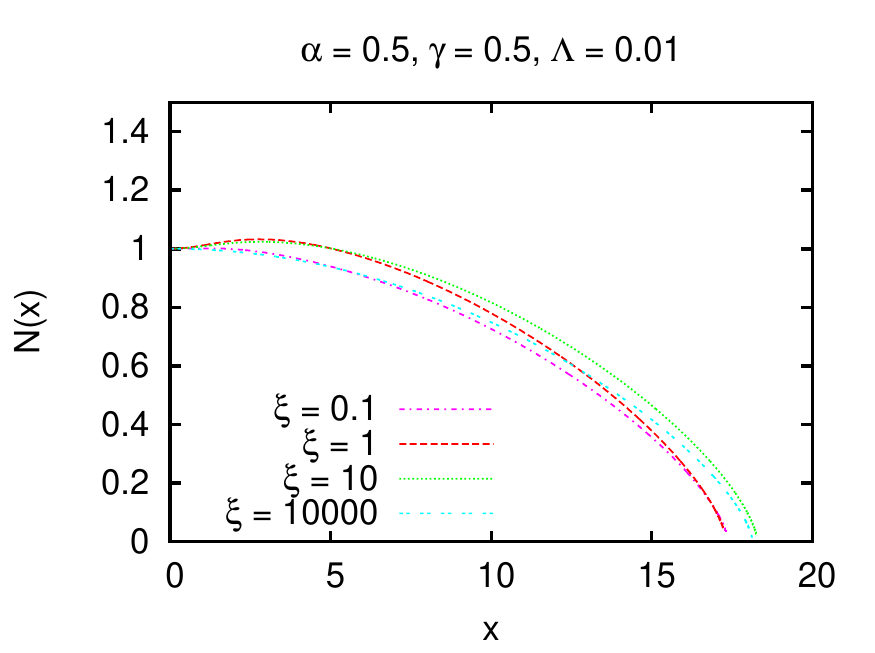}
\end{tabular}
\caption{The metric $N(x)$ function for the Starobinsky model. The zero of $N(x)$ indicates the cosmic horizon $x_{ch}$. For low values of the parameter $\xi$, $x_{ch}$ is lower as $\gamma$ increases, but for large values of the parameter $\xi$, the cosmic horizon is independent of the parameter $\gamma$.}
\label{functionNforStarobinsky}
\end{figure}

\subsection{$f(R) = R + \xi R^2$}

To study the cosmic horizon, $x_{ch}$, we must consider the metric function $N(x)$, since $x_{ch}$ is given by the first zero of $N(x)$. We worked out the field equations for several sets of parameters, and in Figure \ref{functionNforStarobinsky} it is plotted six of those. 

On the top left, the $N(x)$ function is plotted for $\gamma = 0$ and several values of the parameter $\xi$. Since no difference can be observed, we can conclude that $x_{ch}$ is the same for general relativity and in the Starobinsky model, as long as we stay in vacuum. We will denote this cosmic horizon in vacuum by $x_{ch}^0$. In the following plots, we can note that $x_{ch}$ become smaller as $\gamma$ gets bigger. This same feature can be observed in general relativity, and the critical difference to the Starobinsky model appears as the parameter $\xi$ increases: The metric $N(x)$ function approaches the same profile as in the vacuum case, and $x_{ch} \rightarrow x_{ch}^0$. The correction to general relativity acts to decrease the influence of the matter fields on the geometry and do not backreact on this aspect of the geometry. As long as $\xi$ achieves some minimal value,  any value for the parameter $\gamma$ will result in the same cosmic horizon. The graphs in Figure (\ref{functionNforStarobinsky}) just reproduces the same behaviour. 

\subsection{$f(R) = R + \xi R^2 + \chi R^3$}

For the extended Starobinsky model we studied in details only the $R^3$ term in the power series, but it is expected the same qualitative behaviour for the whole power series. This case is not so trivial as the original Starobinsky model, since the $R^3$ term of the action alter the cosmological constant term, and so influences on the value of the cosmic horizon. 

\begin{figure}[htb]
\centering
\begin{tabular}{@{}cc@{}}
\includegraphics[scale=0.8]{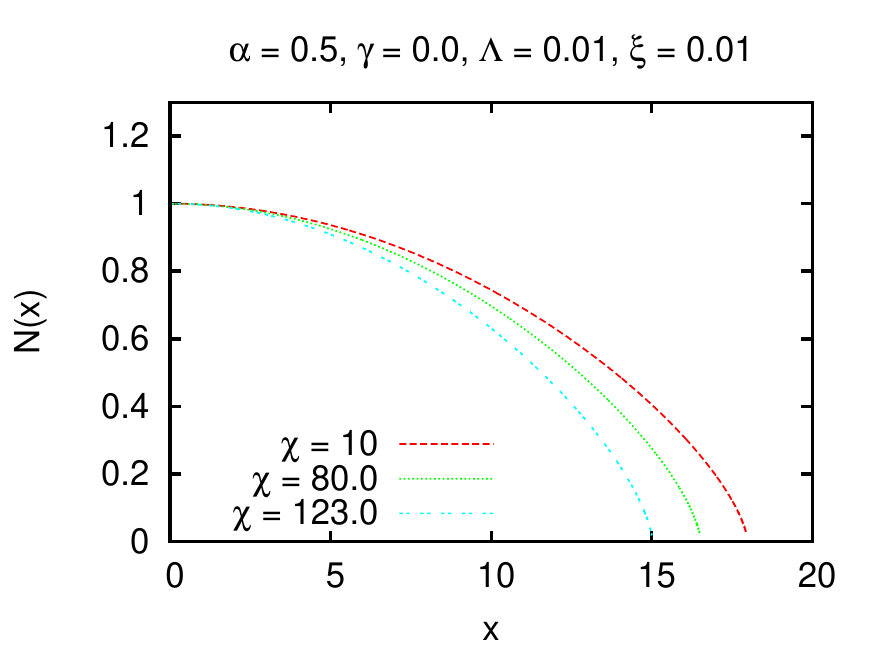} &
\includegraphics[scale=0.8] {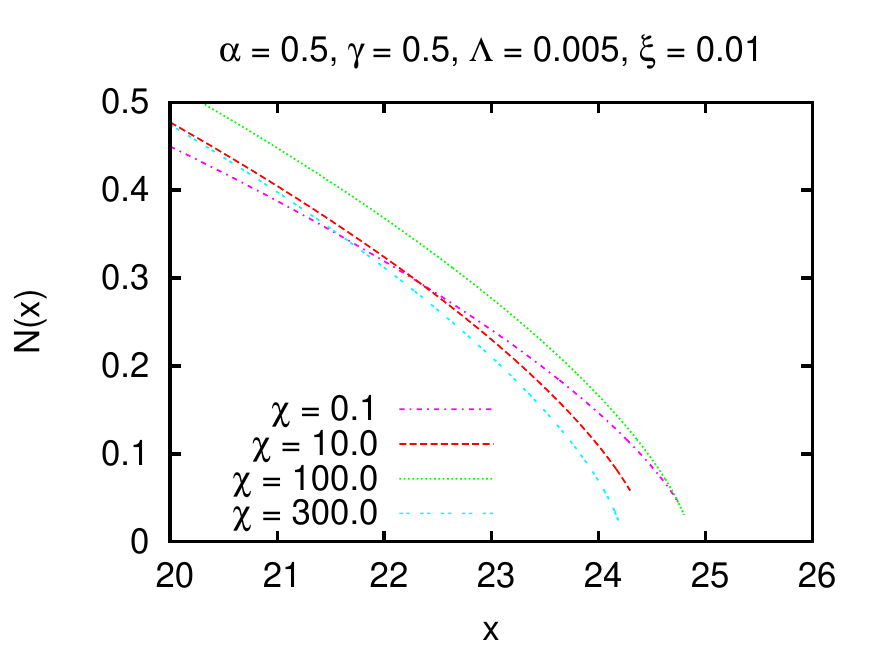} \\
\includegraphics[scale=0.8]{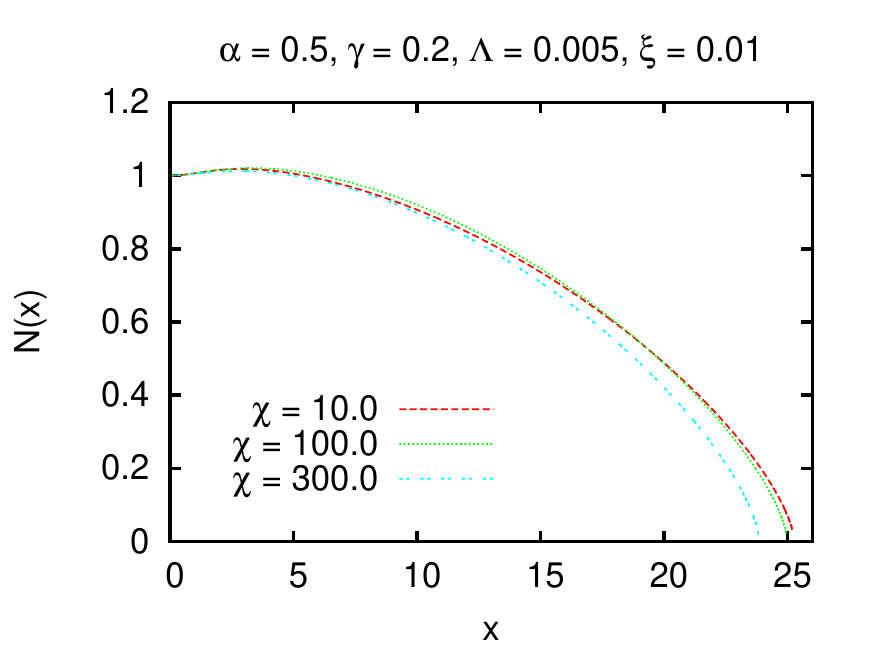} &
\includegraphics[scale=0.8] {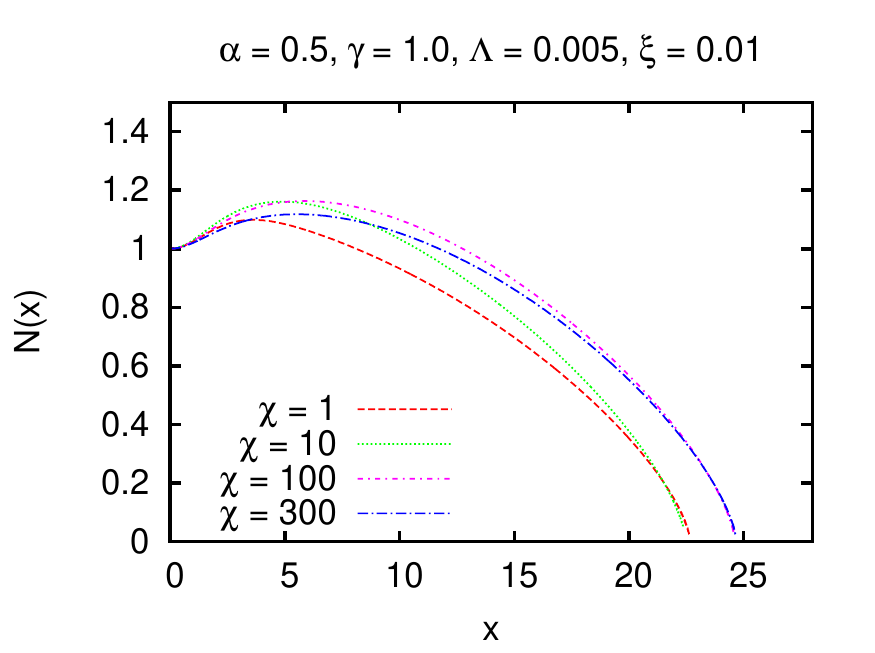} \\
\includegraphics[scale=0.8]{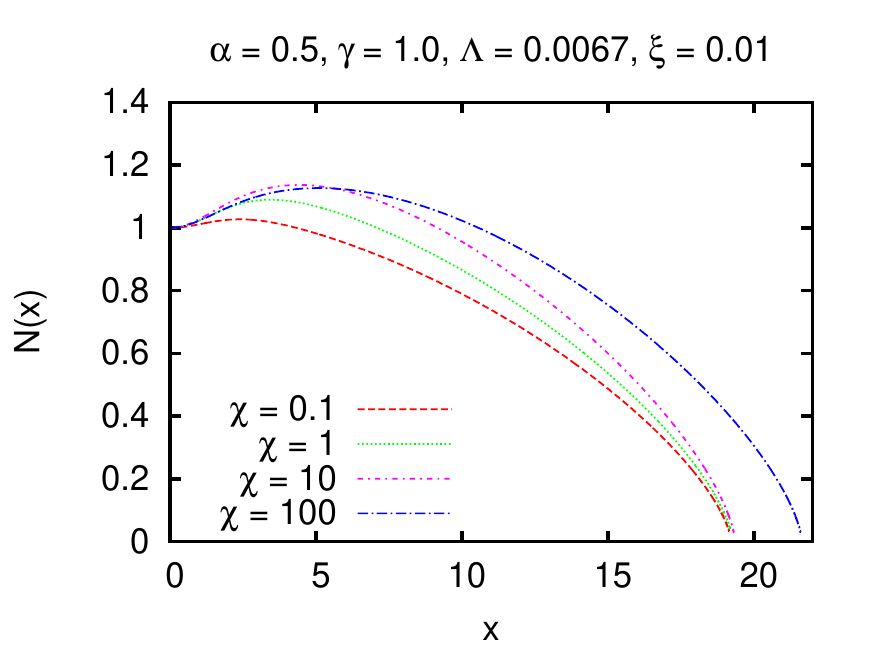} &
\includegraphics[scale=0.8] {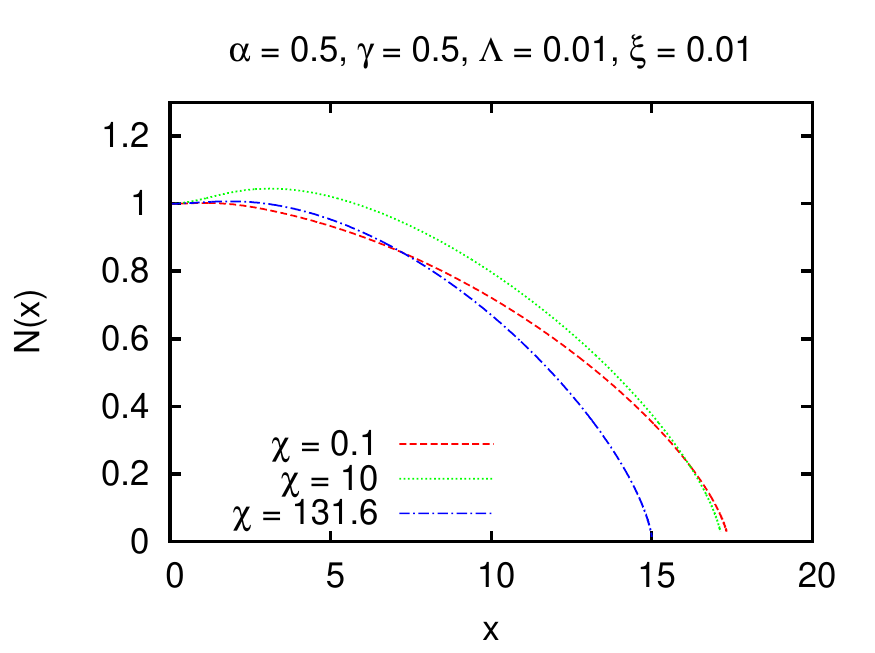}
\end{tabular}
\caption{The metric $N(x)$ function for the extended Starobinsky model with $m=3$. The zero of $N(x)$ indicates the cosmic horizon $x_{ch}$. As the parameter $\chi$ grows, $x_{ch}$ can increase or decrease. It will decrease if the parameter $\gamma$ is small, but will increase if $\gamma$ is big.}
\label{NFunctionExtended}
\end{figure}

In Figure (\ref{NFunctionExtended}) is plotted six graphs for the metric $N(x)$ function. The top left plot was calculated for $\gamma = 0$ and, differently from the pure Starobinsky term, the cosmic horizon is changed as we increase the value of the parameter $\chi$. Let us remember that, asymptotically, the Ricci scalar is given

\begin{equation}
R(\infty) = 4\Lambda + \frac{\chi 64 \Lambda^3}{1 - \chi 48 \Lambda^2},
\end{equation}
which means that the geometry will be stronger or, in other terms, that the cosmological constant will be bigger. As the value for $x_{ch}$ depends on the cosmological constant, it is natural that it changes according to the value of $\chi$. 

The following two plots show a similar behaviour. As we increase the parameter $\chi$, the cosmic horizon become smaller. However, the difference is more pronounced as $\gamma \rightarrow 0$. In the fourth graph, the response of the metric function in respect to $\chi$ invert and the cosmic horizon starts to grow. The explanation for this inversion is in the fact that now we are dealing with two forces. As we increase the parameter $\chi$, we are decreasing the effect of the matter fields in the geometry, just like in the original Starobinsky model, and the cosmic horizon will grow up. But now we are also increasing the value of the cosmological constant, and the effect will be to decrease the value of the cosmic horizon. The relation between this two different effects will be related mostly to the parameter $\gamma$. For low $\gamma$, the matter fields do no strongly influence the geometry, and the effect on the cosmological horizon will be predominant. The consequence is that the cosmic horizon will shrink. For a higher $\gamma$, the influence of the matter on the geometry is big, and the major effect of the correction term will be to diminish this influence. The consequence is that the cosmic horizon will grown.

It is expected that, for a large enough value for the parameter $\chi$, the increase of the cosmological horizon will dominate over the diminishing of the coupling with the matter fields. The precise value where these transitions happens will also depend on the value of $\Lambda$, since the scalar curvature is proportional to it. 

\section{Conclusions}

In this paper we have studied the Abelian Higgs cosmic string in the extended Starobinsky model of gravity, with and without a cosmological constant term. Such a model can be recast as an $f(R)$ theory, where $f(R) = (R - 2\Lambda) + \xi R^2 + \chi R^m$. The parameters $\xi$ and $\chi$ are adimensional, and general relativity is obtained when both goes to zero. The field equations were derived and solved numerically for several values of the parameters $\xi$, $\chi$ and $\Lambda$. We have worked out the powers $m = 3, 4, 8$ and $16$, but our focus was the powers $m = 3$ and $m = 4$. 

The obtained results were compared with the same system in general relativity. For the asymptotically flat case, our main interest was to study the behaviour of the angular deficit generated by the string, since its value can impose an upper constraint in the vacuum expected value for the scalar field. We found that the extended Starobinsky model allows a large upper bound than Einstein's gravity, since as we increase the parameter $\xi$, the angular deficit become smaller. However, the effect of these extra terms, $R^m$, with $m > 2$, are not as strong as the $R^2$ term, and the effect decreases as we increase the value of $m$. We can then conclude that, for this particular regime, only the first terms in a power expansion are relevant in an effective theory of gravity.

For the asymptotically de Sitter case we also compared our obtained results with the ones obtained in general relativity \cite{BezerradeMello:2003ei}. Our interest was to study the cosmic horizon, since it can also impose an upper bound in the vacuum expectation value. For the original Starobinsky model, we have found that as we increase the value of the parameter $\xi$, the cosmic horizon size approaches the same value as in the vacuum case, where the cosmic string decouples from gravity. In a regime with a large parameter $\xi$, the value for the VeV in not relevant for the size of the cosmic horizon. For the extended Starobinsky model we have found another result. The main reason for this different behaviour is due to the fact that the $R^m$ term in the action, with $m > 2$ is able to affect the boundary condition for the scalar curvature. This can be seen as a modification on the cosmological constant. Because of this new effect, as we increase the parameter $\chi$ the cosmic horizon does change, but does not necessary approaches a non-singular asymptotically value. For a small VeV, the main effect of the extended Starobinsky model will be to decrease the cosmic horizon, since the contribution to enlarge the cosmological constant will dominate. For a large VeV, the main effect of the same model, at least for $\chi \approx 10^3$, will be to increase the cosmic horizon, since the contribution to dilute the gravitational effect due to the cosmic string will dominate over the increasing in the cosmological constant.

In any case, if we believe that the extended Starobinsky model can represent an effective theory for gravity, the upper bounds on the vacuum expectation value for the scalar field due to the (possible) existence of cosmic string, should be relaxed when compared with what would be the same bounds in general relativity.

\ack J.P.M.G. was supported by CAPES Fellowship. V.B.B. would like to thank Conselho Nacional de Desenvolvimento Cient\'{i}fico e Tecnol\'{o}gico (CNPq) for partial financial support.

%
%

%
%
\end{document}